\documentclass[preprint,showpacs,showkeys,superscriptaddress]{revtex4}
\usepackage{graphicx}
\usepackage{amsmath, amsthm, amsfonts,amssymb}

\begin{document}

\title{A Multifractal Analysis of Asian Foreign Exchange Markets}

\author{Gabjin Oh}
\affiliation{Pohang Mathematics Institute, Pohang University of Science and Technology, Pohang, Gyeongbuk, 790-784, Korea}

\author{Cheoljun Eom}
\affiliation{Division of Business Administration, Pusan National University, Busan 609-735, Korea}

\author{Shlomo Havlin}
\affiliation{Center for Polymer Studies and Department of Physics, Boston university, Boston, MA 02215} \affiliation{Minerva Center and Department of Physics, Bar-Ilan University, Ramat-Gan 52900,
Israel}

\author{Woo-Sung Jung}
\affiliation{Center for Polymer Studies and Department of Physics, Boston university, Boston, MA 02215}
\affiliation{Department of Physics and Basic Science Research Institute, Pohang University of Science and Technology, Pohang, Gyeongbuk, 790-784, Korea}

\author{Fengzhong Wang}
\affiliation{Center for Polymer Studies and Department of Physics, Boston university, Boston, MA 02215}

\author{H. Eugene Stanley}
\affiliation{Center for Polymer Studies and Department of Physics, Boston university, Boston, MA 02215}

\author{Seunghwan Kim}
\affiliation{Department of Physics and Basic Science Research Institute, Pohang University of Science and Technology, Pohang, Gyeongbuk, 790-784, Korea}
\affiliation{Asia Pacific Center for Theoretical Physics, Pohang, Gyeongbuk, 790-784, Korea}

\date{\today}

\begin{abstract}
We analyze the multifractal spectra of daily foreign exchange rates for Japan, Hong-Kong, Korea, and Thailand with respect to the United States Dollar from 1991 to 2005. We find that the return time series show multifractal spectrum features for all four cases. To observe the effect of the Asian currency crisis, we also estimate the multifractal spectra of limited series before and after the crisis. We find that the Korean and Thai foreign exchange markets experienced a significant increase in multifractality compared to Hong-Kong and Japan. We also show that the multifractality is stronge related to the presence of high values of returns in the series.

\end{abstract}

 \maketitle

\section{Introduction}
Economic systems are widely acknowledged as extremely complex, and
have recently become an interesting area of study for physicists
as well as economists \cite{Mantegna&Bouchaud}. Many previous
studies have found that time series of financial markets exhibit
some non-linear properties developed in statistical physics
\cite{Mantegna(a), {Oh}}. The prices in financial markets are
created by non-trivial interactions among heterogeneity agents and
complex events occurring in the external environment. In other
words, both micro and macro variables with various time scales are
involved in the pricing mechanism.

The properties observed in financial time series include
long-memory in volatility \cite{Oh}, a multifractal nature
\cite{Muzy, Sornette, Plamen, Parisi, Mandelbrot}, and fat tails
\cite{Mantegna(a)} among others; these are sometimes referred to
as the {\em stylized facts}. The multifractal concept, which is
now well developed in the fields of statistical physics and
nonlinear dynamics, is a well-known feature of complex systems
\cite{Plamen, Parisi, Mandelbrot, Matia, Kwapie}. Multifractality
has been discovered in systems as diverse as earthquakes
\cite{Parisi}, turbulence systems, biological time series
\cite{Plamen}, as well as financial markets \cite{Muzy, Kwapie}.

Previous studies have found evidence for a relationship between
the complexity of   a system and its degree of multifractality
\cite{Plamen}. For example, the degree of multifractality in data
generated by a multiplicative cascading process is directly
related to the long-range correlations of the magnitude time
series \cite{Ashkenazy}. Other factors that could affect the
multifractality of time series include time correlations and the
probability distribution of the data \cite{Matia, Kwapie}.
However, it is still not clear what is the origin of
multifractality in financial markets.

In this paper we study the multifractal properties of a financial
time series: the daily return of four foreign exchange (FX)
markets. We consider Japan (JPY/USD), Hong-Kong (HKD/USD), Korea
(KRW/USD), and Thailand (THD/USD) from 1991 to 2005. We employ
multifractal detrended fluctuation analysis (MF-DFA)
\cite{Kantelhardt} to measure the nonlinear features of the time
series, in particular their multifractal spectra. To test their
significance we randomly shuffle the series to remove any temporal
correlations, and find that these spectra narrow significantly. In
other words, we find that temporal correlation plays an important
rule in the multifractality of the data, similar to Matia et al.
\cite{Matia}.

To detect changes in market complexity before and after the Asian
currency crisis, we divide the series into two periods before and
after the crash and calculate their multifractal spectra
separately. We find that for Korean and Thailand the degree of
multifractality increased significantly after the Asian currency
crisis, while the FX markets of Hong Kong and Japan did not. We
therefore suggest that both Korea and Thailand have been more
influenced by the Asian currency crisis. We also examine the
effect of return values above a certain threshold in the FX
markets on the market complexity. We find that for all countries,
market complexity is related to higher returns.


In the next Section, we describe the financial data and our
methodology. In Section 3, we present our results of this study.
Section 4 concludes the article.

\section{Data and Methodology}

We investigate the multifractal properties of Asian FX markets
(returns to U.S. Dollars) from 1991 to 2005 for four countries:
Japan (JPY/USD), Hong-Kong (HKD/USD), Korea (KRW/USD), and
Thailand (THB/USD). The data are obtained from
http://www.federal/reserve.gov/RELEASES/. In all data sets used in
this paper, we remove the year 1997 to eliminate any abnormalities
due to the market crash itself. The return time series is
calculated by the log-difference of daily prices: ${r(t)} =
\ln{P(t)}- \ln{P(t-1)}$, where $P(t)$ is the foreign exchange rate
on day $t$. We divide the whole series into two sub-periods: DATA
A from 1991 to 1996 (before the crisis) and DATA B from 1998 to
2005 (after the crisis). This allows us to study the influence of
the Asian currency crisis on market complexity. We employ the
multifractal detrended fluctuation analysis (MF-DFA) method to
determine the multifractal properties of the time series. The
MF-DFA method was proposed by Kantelhardt et al.
\cite{Kantelhardt}, and can be explained by the following three
steps.

Step (1): We subtract the average value of the time series from
each point $x(i)(\equiv r(t))$, then accumulate the series:

\begin{equation}\label{e1}
y(i)= \sum_{k=1}^{i} [x(k) - \bar{x}],
\end{equation}
where $x(i)$ is the $i^{\mathrm th}$ point and $\bar{x}$ is the
mean of all $\{x(k)\}$. This step represents the original data as
an accumulated profile, $y(i)$.

Step (2): The profile $y(i)$ is divided into $N_s$ boxes of length
$s$. In each box $v (1\leq v \leq N_{s})$ , the trend is estimated
by an $m$-order polynomial using the least-squares method. The
best-fit curve of a given box is expressed as $y_{v}(i)$. By
subtracting $y_{v}(i)$ from $y(i)$, possible trends are removed
\cite{Peng}. This process is applied to every box, and the
fluctuations in that box are then calculated as

\begin{equation}\label{e2}
F_{2}(s,v) \equiv \frac{1}{s} \sum_{i=1}^{s}(|y((v-1)s+i) -
y_{v}(i)|)^2.
\end{equation}

Step (3): We compute the mean $q$-order moment $F_{q}(s)$ of the
series by averaging the appropriate function of $F_2$ over all
boxes. In this way we obtain a scaling relation with box size $s$:
\begin{equation}\label{e3}
F_{q}(s) \equiv \{\frac{1}{N_{s}} \sum_{v=1}^{N_s}
F_{2}(s,v)^{q/2}\}^{1/q} \sim s^{h(q)}.
\end{equation}

The exponent $h(q)$ depends on $q$. In general, the multifractal
(MF) scaling exponent $\tau(q)$ is related to $h(q)$ through

\begin{equation}\label{e4}
\tau(q) = qh(q)-D_{f},
\end{equation}
where $D_{f}$ is the fractal dimension of a geometric object. In
our case, $D_{f} = 1$. The MF exponent $\tau(q)$ represents the
temporal structure of the time series as a function of the various
moments $q$. That is, $\tau$ reflects the scale-dependence of {\em
smaller} fluctuations for {\em negative} values of $q$, and larger
fluctuations for positive values of $q$. In the special case that
$\tau(q) = \alpha q$ is a linear function, the time series can be
regarded as a monofractal and $\alpha$ is the singularity strength
or H\"{o}lder exponent. If $\tau(q)$ increases nonlinear with $q$,
then the series is multifractal. In this case we can calculate the
MF spectrum $f(\alpha)$ by a Legendre transform of $\tau(q)$, as
defined by

\begin{equation}\label{e5}
f(\alpha) \equiv \alpha q - \tau(q), ~~ \alpha \equiv \frac
{d\tau(q)}{dq},
\end{equation}
where $f(\alpha)$ is the dimension of the time series. If the time
series is monofractal, $f(\alpha)$ is a delta function, there is
only one value of $\alpha$; otherwise, there is a distribution of
$\alpha$ values.

\section{Results}
\label{sec:RESULTS}

We have analyzed the multifractal spectra of Asian FX markets
using the above MF-DFA method. The pricing mechanisms may well be
complex, due to the Asian currency crisis in late 1997 as well as
due to various internal and external events. The Asian currency
crisis had an impact on almost all Asian FX markets \cite{Gabjin}.
Here, we study the multifractal properties of four markets and try
to identify how the crisis may have affected their
multifractality.

In Fig. 1, we show the return time series of Hong-Kong (a), Japan
(b), Korea (c), and Thailand (d). The Korean and Thai FX markets
clearly have higher volatility after the Asian currency crisis in
1997, but the Japanese and Hong-Kong markets show no obvious
change.

We now describe the multifractal properties of the four markets.
The results of MF-DFA analysis are presented in Fig. 2. To test
how significant is the multifractality, we also perform the
analysis on shuffled time series created by randomly shuffling the
data. Figs. 2(a) and (b) show the fluctuation spectra $F_{q}(s)^q$
of the original and shuffled HKD/USD series respectively. These
logarithmic plots indicate that in both cases $F_{q}(s)^q$ is a
power-law with an exponents depending on $q$. Fig. 2(c) and (d)
display the multifractal scaling function $\tau(q)$ of the
original and shuffled data. We calculated $\tau(q)$ from the
power-law relation between $F_{q}(s)^q$ and $s$, using scales in
the range $40<s<600$. This is since below $s=40$ there is
discreteness effects (original and shuffled). We find that the two
datasets behave similarly almost linear with q for negative
moments, but show significant non-linearities for positive
moments. This means that the larger fluctuations have changed
dramatically in the shuffled series.

To explicitly observe the multifractality we can convert $q$ and
$\tau(q)$ to $\alpha$ and $f(\alpha)$ by a Legendre transform.
Fig.~2(e) shows the multifractal spectra $f(\alpha)$ of the
original market series. We find that the singularity strengths
$\alpha$ of the markets lie within the following ranges: $-0.04
\leq \alpha_{\rm Hong-Kong} \leq 1.10$, $0.34 \leq \alpha_{\rm
Japan} \leq 0.63$, $0.29 \leq \alpha_{\rm Korea} \leq 0.85$, and
$0.08 \leq \alpha_{\rm Thailand} \leq 0.92$. Fig.~2(f) presents
the same information for the shuffled time series: $0.18 \leq
\alpha_{\rm Hong-Kong(shuffle)} \leq 0.73$, $0.46 \leq \alpha_{\rm
Japan(shuffle)} \leq 0.61$, $0.43 \leq \alpha_{\rm Korea(shuffle)}
\leq 0.64$, and $0.39 \leq \alpha_{\rm Thailand(shuffle)} \leq
0.67$. Figs.~2(e) and (f) show that the multifractal spectra
$f(\alpha)$ are narrower for the surrogate time series, from which
all temporal correlations have been removed. Our results indicate
that the temporal fluctuations in Asian FX markets show signature
of multifractality.

The Asian currency crisis had an influence on almost all the Asian
FX markets, so it is reasonable to assume that their status may
have changed significantly. An interesting question is how the
Asian currency crisis has influenced the market complexity. We
divided each time series into two sub-periods: (DATA A) and (DATA
B) before and after the crisis respectively. Fig. 3 shows the
multifractal spectra $f(\alpha)$ of the original and surrogate
time series, which remove the nonlinearity from the original data
\cite{Theiler}, of both periods for all four FX markets. We find
that the multifractal spectra of all the surrogate data set is
reduced significantly than those of the original data. The Hong
Kong and Japanese FX markets show a similar degree of
multifractality before and after the crisis, while the Korean and
Thai markets change significantly and the multifractal spectra
become broader. In other words, the complexity of the Korean and
Thai FX markets has been increased after the Asian currency
crisis. Fig. 3(e) and (f) shows the change in the degree of
multifractality $\Delta \alpha$ of both the original and surrogate
data for each market. We conjecture that since the Japanese FX
market is the most mature, it was also the least influenced by the
Asian currency crisis. The emerging markets of Korea and Thailand,
on the other hand, were greatly influenced by the crash. As for
the Hong Kong FX market, since Hong Kong chose to have a fixed
exchange rate with the U.S. dollar (called the Peg system) both
periods have similar broad spectra. The Asian currency crisis thus
increased the complexity of the Korean and Thai FX markets,
perhaps because its aftermath spurred the development of new
government policies in those countries.

Table 1 shows the degree of multifractality $\Delta \alpha$ of the
original, shuffled data and before and after the crisis for all
countries used in this paper. This quantity shows that for the
original data the multifractal spectra have more broader than
those of shuffled data and the degree of multifractality both the
Korean and Thailand FX markets increases significantly after the
crisis.

We have observed that temporal correlations are not linear but
posses multifractality in the time series. It is interesting to
note that the shuffled time series removed the time correlation
still show some multifractality. It is widely accepted that the
distribution of returns in a financial market follows a power law,
with an exponent close to 3 \cite{Mantegna(a)}. In other words,
there are many higher values that cannot be predicted by the
pricing mechanism of the efficiency market hypothesis (EMH)
\cite{Fama}, which is also widely used in the financial
literature. We will now investigate the influence of high returns
on the multifractality. To do this, we generate a multifractal
noise data set using the wavelet-cascade model introduced by
Arneodo {\em et. al.} \cite{Arneodo}. We employ the log-normal
random variable $w$ to generate the multifractal noise data. In
this case, the degree of multifractality of the created data is
determined by the parameters such as the mean, $\mu$, and the
standard deviation, $\sigma$, of $\ln(w)$ and it is positively
related to the $\sigma$ value. Where $\ln w$ is a coefficient of
the normal distribution with $\mu$ and $\sigma$. We can create the
artificial data that have the different degree of multifractality
with respective to the mean, $\mu$, and the standard deviation,
$\sigma$. The multifractal spectrum is given by $f(\alpha) =
\frac{-(\alpha + \mu/\ln2)}{2\sigma^2} \ln2 +1$ and the
multifractal spectrum width as the degree of multifractality is
$\frac{2 \sqrt{2} \sigma}{\sqrt{\ln 2}}$.

 To verify the
relationship between the degree of multifractality and the extreme
values, we creates 100 data sets with $2^{17}$ data points and
calculate the tail exponent, $\gamma$, of power law distribution,
$p(x) \sim x^{\gamma}$ using method proposed by A. Clauset {\em
et. al.} \cite{Clauset}. Fig. 4 shows the relationship between the
degree of multifractality and the power-law exponents using the
multifractal noise data sets created with $\mu=\{-0.6 \times \ln
2, -0.7 \times \ln2, -0.8 \times \ln2, -0.9 \times \ln2, -1 \times
\ln2 \} $ and various $\sigma$ values in the ranges from 0.2 to
0.4. The circles (red), sqaures (green), diamons (blue), triangles
(pink), and stars(black) correspond to the multifractal noise data
created with $\mu= \{-0.6\times \ln2, -0.7\times \ln2, -0.8\times
\ln2, -0.9\times \ln2$, $-1 \times \ln2$ \}, respectively. In Fig.
4, we observe that regardless of $\mu$, the exponent, $\gamma$
increases as the degree of multifractality increases. In other
words, the degree of multifractality has strongly relation to the
existing of the extreme values in the multifractal noise data.

To verify the result observed in the Fig. 4 to the foreign
exchange markets, we create a new version of each time series by
eliminating values above a certain threshold, $T$ in units of the
standard deviation of the time series. The eliminated data points
are replaced by linear interpolation. As the threshold $T$
increases, the time series will retain higher values.

Fig. 5(a), (b), (c), and (d) display the dependence of the degree
of multifractality $\Delta\alpha$, defined by the range of
singularity strengths $\alpha$, on the threshold $T$ for original
and surrogate data. The open and filled circles indicate the
original and surrogate data, respectively. We find that in all
four countries, FX market complexity is related to the presence of
very high return values. However, for the surrogate data removed
the nonlinearity from the original data is an independent from the
threshold T. Market values are created by non-trivial interactions
between heterogeneity agents and the influence of internal and
external events. The results of Fig. 5 seem to indicate that more
complex markets are more likely to produce high returns.


\section{Conclusions}
\label{sec:CONCLUSIONS}

We investigated the properties of time series from four Asian
foreign exchange markets, and found two factors affecting their
multifractality as measured by the MF-DFA method. First, we found
that temporal correlations in the data contribute to the
multifractality of all four FX markets. Second, we find that
market complexity and multifractality are positively related to
the presence of high return values in the series. Further studies
will examine both aspects of FX markets more extensively.

To observe how the Asian currency crisis influenced these FX
markets, we estimated their multifractal properties both before
and after the crisis. We found that in both Korea and Thailand,
the degree of multifractality in the FX market significantly
increased after the Asian currency crisis. Japan and Hong Kong,
however, were almost unaffected. We argue that the market crash
affected Korea and Thailand more strongly because they are typical
emerging markets; these countries probably introduced new policies
(and thus additional complexities) to help control the aftermath.
Japan's mature market was little changed by the crisis, however,
and Hong Kong uses a fixed exchange rate.


\begin{acknowledgments}
This work was supported partially by the Korea Research Foundation
Grant funded by the Korean Government (KRF-2007-412-J02303) and
partially by the KOSEF under the grant number
R01-2008-000-21065-0.
\end{acknowledgments}

\begin{figure}[tb]

\includegraphics[height=16cm, width=16cm]{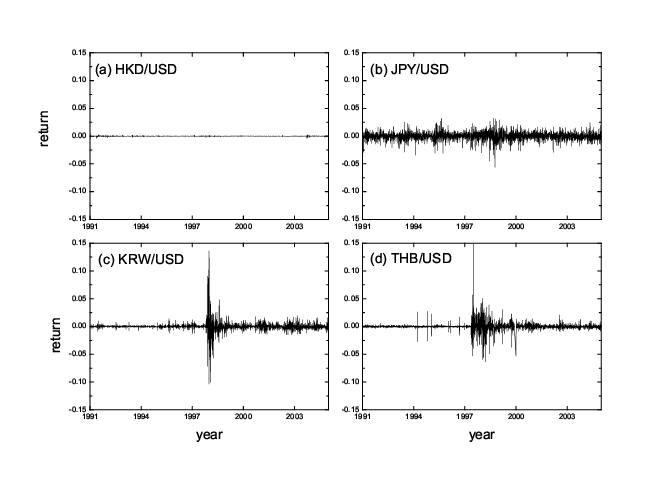}

\caption[0]{Returns $x(i)$ of four Asian foreign exchange markets:
Hong-Kong (a), Japan (b), Korea (c), and Thailand (d).}

\end{figure}

\begin{figure}[tb]

\includegraphics[height=17cm, width=16cm]{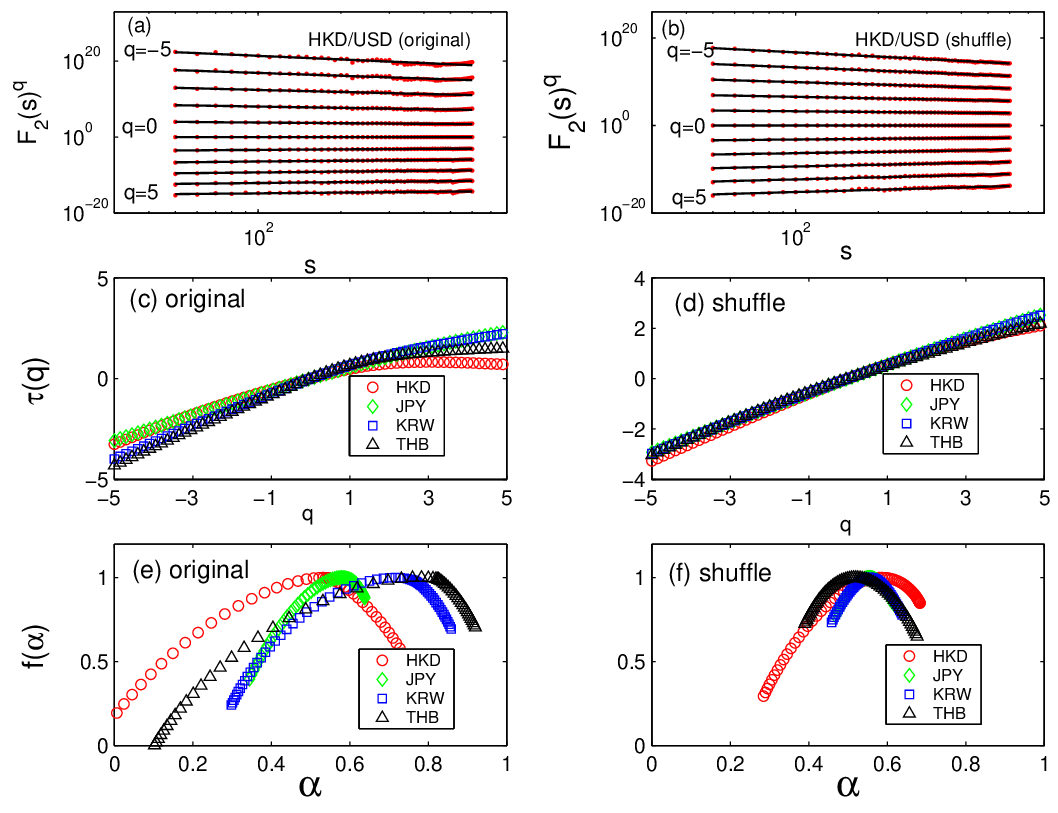}

\caption[0]{Panels (a) and (b) show the scaling function
$F_q(s)^q$ for various moments $q$ of the Hong-Kong FX market, for
the original (Fig. 1(b)) and surrogate (shuffled) time series
respectively. Panels (c) and (d) show the multifractal scaling
functions $\tau(q)$ of the return and surrogate time series for
all four foreign exchange markets. Panels (e) and (f) show the
fractal dimension $f(\alpha)$ obtained by a Legendre
transformation of (c) and (d) respectively. Red circles refer to
Hong Kong (HKD), green diamonds to Japan (JPY), blue squares to
Korea (KRW), and black triangles to Thailand (THB).}

 \label{fig2}
\end{figure}

\begin{figure}[tb]

\includegraphics[height=18cm, width=16cm]{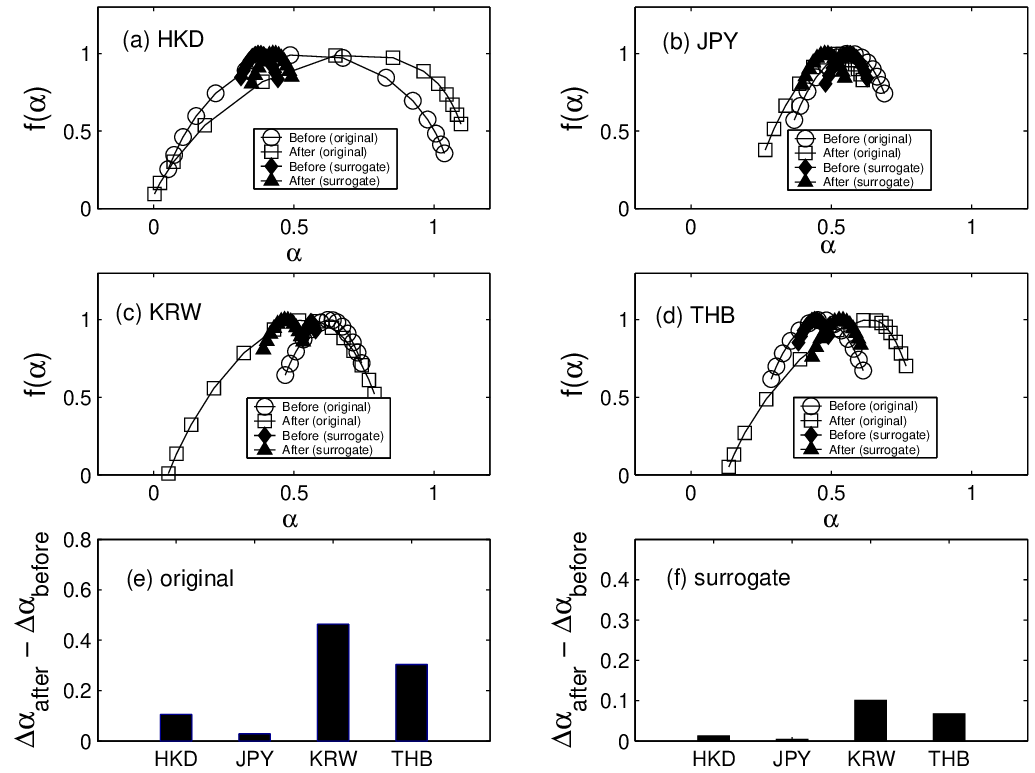}

\caption[0]{The multifractality spectrum $f(\alpha)$, before
(circles) and after (squares) the Asian currency crisis. The four
panels refer to Hong Kong (a), Japan (b), Korea (c), and Thailand
(d). Panel (e) and (f) plots the difference in the range of
multifractality $\Delta \alpha$ for the original and surrogate
data.}
\end{figure}

\begin{figure}[tb]

\includegraphics[height=16cm, width=16cm]{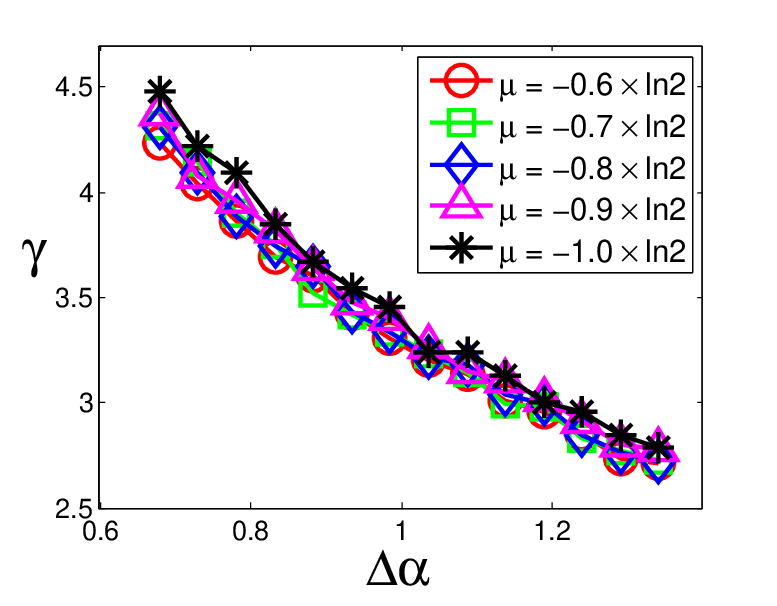}

\caption[0]{The relationship between the degree of multifractality
and the tail exponent of power law distribution. We generate 100
 multifractal noise data sets with $2^{17}$ data points with respect
to $\mu=\{-0.6 \times \ln2, -0.7 \times \ln2, -0.8 \times \ln2,
-0.9 \times \ln2, -1 \times \ln2 \} $ and various $\sigma$ values
in the ranges from 0.2 to 0.4 and calculate the exponent of
power-law distribution function. The circles (red), sqaures
(green), diamons (blue), triangles (pink), and stars(black)
correspond to the multifractal noise data created with $\mu=
\{-0.6\times \ln2, -0.7\times \ln2, -0.8\times \ln2, -0.9\times
\ln2$, $-1 \times \ln2$ \}, respectively.}

\end{figure}

\begin{figure}[tb]

\includegraphics[height=16cm, width=16cm]{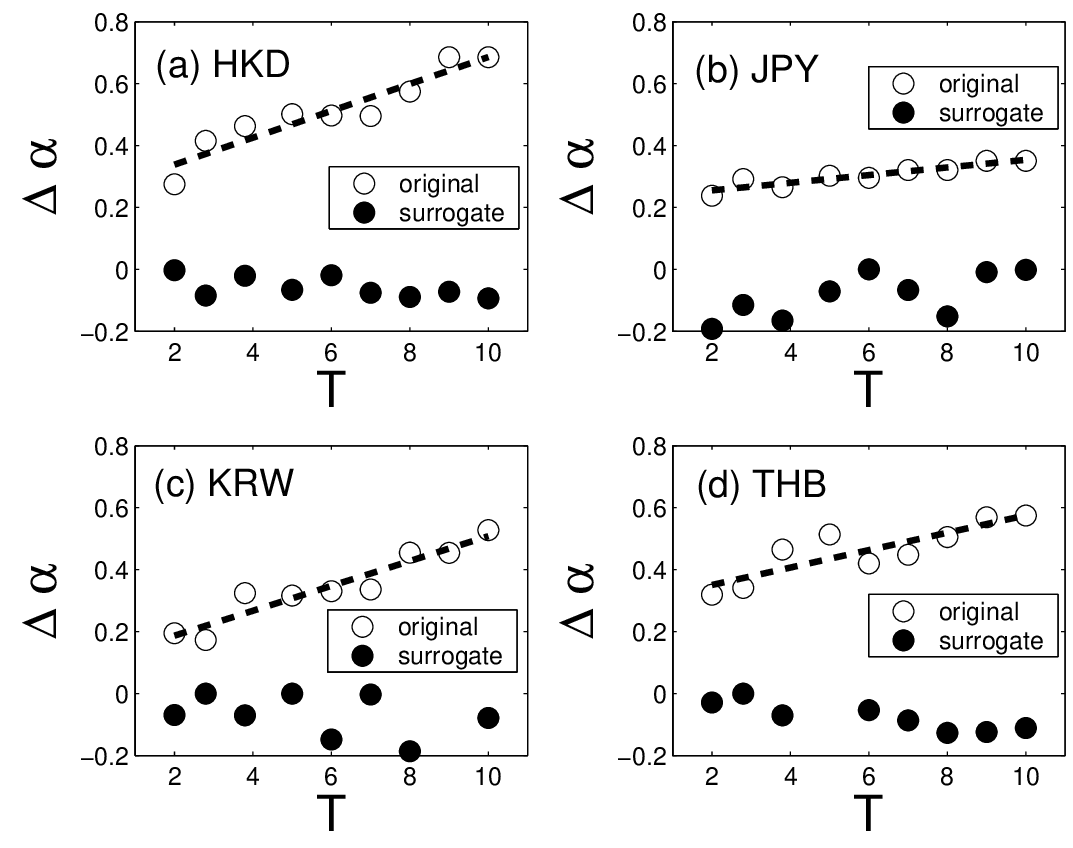}

\caption[0]{The multifractality of time series from which returns
above a threshold $T\sigma$ have been deleted and interpolated.
The panels refer to Hong-Kong (a), Japan (b), Korea (c), and
Thailand (d). The open and fill circles indicate the original and
surrogate time series. }
\end{figure}









\newpage


\begin{table}[ht]
\caption{The degree of multifractality $\Delta \alpha$ of all the
countries}
\begin{tabular}{c c c c c}
\hline \hline Country ~~ & ~~Original ($\Delta\alpha$) ~~& ~~Shuffled ($\Delta\alpha$) ~~& ~~Before ($\Delta\alpha$)~~& ~~After ($\Delta\alpha$)\\
\hline \\
HongKong &1.14 &0.55 &0.98 & 1.09 \\
Japan  &0.29 &0.15&0.32&0.34 \\
Korea  &0.55 &0.20&0.27&0.73 \\
Thailand  &0.83 &0.28&0.32&0.63 \\ \hline
\end{tabular}
\end{table}

\end{document}